\documentclass[doublecol]{epl2} 

\usepackage[svgnames]{xcolor}
\usepackage[colorlinks=true,allcolors=black]{hyperref}  
\newcommand{\gsim}{\raisebox{-0.13cm}{~\shortstack{$>$ \\[-0.07cm]
      $\sim$}}~}

\title{Intermittency from instanton calculus at the\\transition to turbulence and fusion rules}
\shorttitle{Intermittency from instantons and fusion rules}

\author{T. Schorlepp\inst{1}\thanks{E-mail: \email{timo.schorlepp@nyu.edu}} \and R. Grauer\inst{2}\thanks{E-mail: \email{grauer@tp1.rub.de}}}
\shortauthor{T. Schorlepp \etal}

\institute{
  \inst{1} Courant Institute of Mathematical Sciences, New York University,
New York, NY, USA\\
  \inst{2} Institute for Theoretical Physics I, Ruhr University Bochum,
Bochum, Germany
}

\abstract{
Understanding intermittency of turbulent systems from the underlying differential equations is an outstanding problem in fluid dynamics. Here, in the example of Burgers turbulence as a stringent test, we introduce a semi-analytical method that yields high-order structure function exponents by combining instanton calculus, fusion rule predictions, and low-order statistical inputs from direct numerical simulations (DNS). We use instanton predictions, calibrated by DNS, to evaluate high velocity gradient (VG) moments at the onset of intermittency, and then infer scaling exponents in fully developed turbulence via fusion rules. We show that the method captures the crossover at $\mathrm{Re}_\lambda \approx 1$ in the VG moment scaling, highlight the necessity of including fluctuations around instantons, and discuss future extensions.
}

\begin{document}

\maketitle


\section{Introduction}
Turbulence is often referred to as the last unsolved problem
in classical physics~\cite{sreenivasan-schumacher:2025}.
Of outstanding importance is understanding the anomalous
scaling of structure functions or, equivalently, the non-self-similar behavior of probability distribution functions (PDFs) of velocity increments, leading from nearly Gaussian shapes at large separations of the increments to heavy-tailed distributions at small distances~\cite{frisch:1995}.
Various theoretical methods have been used to tackle this problem.
One class of methods is represented by phenomenological models~\cite{meneveau-sreenivasan:1991,she-leveque:1994,dubrulle:1994,grauer-krug-marliani:1994, yakhot:2001,yakhot:2006,eling-yaron:2015,apolinario-moriconi-pereira-etal:2020}, 
which use only global properties of the underlying partial differential equations (PDEs) such as the Navier--Stokes equations (NSE), Burgers equation, magnetohydrodynamic (MHD) equations, e.g., regarding a direct energy cascade due to a quadratic nonlinearity of advection type, or properties of dissipation at localized structures.
These models have shown remarkable agreement with experiments or direct numerical simulations (DNS), and have highlighted the importance of accounting for quasi-singularities such as vortex tubes, sheets or shocks~\cite{duchon-robert:2000}.
However, phenomenological models usually have free parameters that are not determined from the underlying PDE, and the geometry of quasi-singular structures is inferred from experiments or DNS.
Another class of methods follows a direct way, i.e., to determine from the underlying PDE the properties of the energy spectrum, structure functions and increment statistics. These methods are extremely diverse and mostly motivated from quantum and statistical field theory.
A distinction can be made between 
perturbative (direct interaction approximation~\cite{kraichnan:1959,wyld:1961}, renormalization group~\cite{yakhot-orszag:1986, antonov-borisenok-girina:1996})
and non-perturbative methods (fusion rules~\cite{lvov-procaccia:1996,benzi-biferale-toschi:1998,schumacher-sreenivasan-yakhot:2007},
replica symmetry breaking~\cite{bouchaud-mezard-parisi:1995},
functional renormalization group~\cite{mathey-gasenzer-pawlowski:2015,canet:2022,fontaine-tarpin-bouchet-etal:2023}, instanton calculus~\cite{gurarie-migdal:1996,falkovich-kolokolov-lebedev-etal:1996,balkovsky-etal:1997,falkovich-lebedev:2011,grafke-grauer-schaefer_2015,apolinario-moriconi-pereira-2019,fuchs-herbert-rolland:2022,burekovic-schaefer-grauer:2024}).
We also mention the access to vorticity statistics via area laws \cite{migdal:1994,migdal:2019,iyer-etal:2021}, and numerical studies of fusion rules and velocity gradient (VG) scaling~\cite{ishihara-kaneda-yokokawa-etal:2007,schumacher-sreenivasan-yakhot:2007,chakraborty-frisch-pauls-ray:2012,yakhot-donzis:2017,friedrich-margzoglou-biferale-etal:2018,gotoh-yang:2022,khurshid-etal:2023,buaria-pumir:2025,buaria-pumir:2026,dey-mukherhee-banerjee-etal:2026}.
As we recall in more detail below, in stochastically driven fluids, an \textit{instanton} represents the most probable field configuration in space and time that leads to a specific extreme event, such as the formation of a steep velocity gradient. By identifying this dominant trajectory and perturbations around it, one can directly probe the extreme tails of intermittent PDFs. In contrast, \textit{fusion rules} establish a link between different flow regimes by connecting scaling behavior of ``fused'', local observables, such as velocity gradients, to multi-point structure functions in the inertial range.
\section{Our approach}
We present a general, non-perturbative method in this letter which computes high-order structure function exponents at large Reynolds numbers (Re) using a semi-analytical prediction scheme, \textit{combining} the instanton approach for VG moments at low to moderate Re with results from fusion rules and DNS. This is conceptually in line with recent works~\cite{yakhot-donzis:2017,gotoh-yang:2022,carbone-wilczek:2024} which use VG information at the onset of intermittency to draw conclusions about flows at asymptotically large Re.
We emphasize that while the presented approach requires considerable technical effort, it provides a physically interpretable route towards understanding turbulence, which marks the main methodological contribution of this letter.
Below, we compute VG PDF tail asymptotics from the governing PDE via the instanton formalism, and combine these theoretical predictions with DNS to calibrate the low-order core fluctuations (such as the second moment), as the latter fall outside the range of applicability of the instanton method. The resulting ``hybrid'' or semi-analytical predictions are then combined with fusion rules to calculate structure functions. To show that our method's predictive power for extreme events is independent of low-order DNS input, we complement our results with an extended self-similarity (ESS) analysis.
We demonstrate the method on the example of Burgers
turbulence~\cite{frisch-bec:2002,bec-khanin:2007}.
Notably, deriving its (well-known) structure function exponents from first principles still poses a formidable challenge for theoretical methods~\cite{mathey-gasenzer-pawlowski:2015,canet:2022} since intermittency is particularly strong and the distributions of velocity increments and VGs strongly deviate from Gaussianity. In another sense, however, Burgers turbulence is simple, since the dominant structures, shocks, are extremely stable and do not allow for symmetry breaking as, e.g., the transition from vortices to sheets in the NSE \cite{schorlepp-grafke-may-grauer:2022}, which would require the inclusion of zero modes~\cite{burekovic-schaefer-grauer:2024,schorlepp-grafke-grauer:2023} in our approach. Hence, Burgers turbulence forms an ideal test bed and rigorous proof-of-concept for the proposed general method.
\section{Outline} 
\begin{figure}
    \centering
    \includegraphics[width = \linewidth]{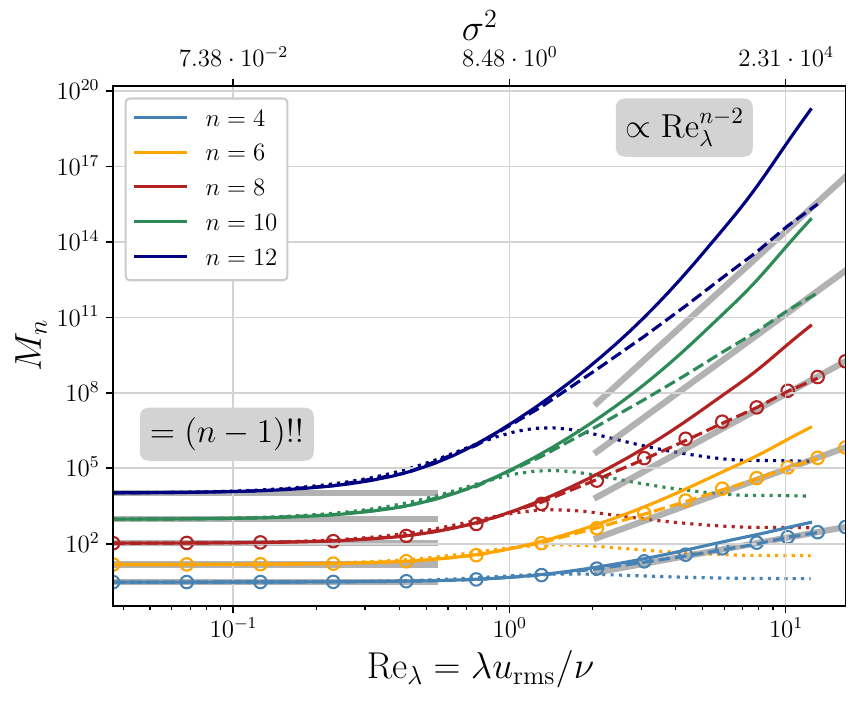}
    \caption{Normalized VG moments $M_n$ (eq.~(\ref{eq:mn-def})) at different Taylor--Reynolds numbers $\mathrm{Re}_\lambda$ from integrating (i) only the instanton contribution $\rho(a) \approx \mathrm{const} \times \exp \left(-S^{\mathrm{I}}(a) / \sigma^2\right)$ to the VG PDF (dotted lines), (ii) the one-loop approximation~(\ref{eq:inst-approx-pdf}) including Gaussian fluctuations around the instanton (solid lines), and (iii) eq.~(\ref{eq:inst-approx-pdf}) for the $n$-th moment in the numerator of eq.~(\ref{eq:mn-def}) with rescaled left PDF tail, and DNS for the second moment in the denominator of eq.~(\ref{eq:mn-def}) (dashed lines). Open circles show the results from DNS of eq.~(\ref{eq:burg}). The thick gray lines indicate the theoretically expected behavior.}
    \label{fig:mn-result}
\end{figure}

We first recall the path integral formulation of Burgers turbulence, which is the basis for approximating statistical quantities using instantons and fluctuations around them~\cite{schorlepp-kormann-luebke-etal:2025}. We then briefly introduce fusion rules for VG moments in Burgers turbulence~\cite{friedrich-margzoglou-biferale-etal:2018}, and present the Re dependence of the normalized VG moments
\begin{equation}
M_n = \frac{\left \langle (\partial_x u)^n \right \rangle}{\left \langle (\partial_x u)^2 \right \rangle^{n/2}}\,,
\label{eq:mn-def}
\end{equation}
and their relation to structure function exponents $\zeta_q$. Using instantons and Gaussian fluctuations around them, the normalized VG moments are calculated accurately at moderate Re, cf.\ fig.~\ref{fig:mn-result}. Combining this with fusion rules, anomalous exponents~$\zeta_q$ at large Re are computed. In the conclusion, we critically summarize our results and give an outlook on further improvements and applications of the method to other systems, such as 3D NSE turbulence.
%
%
%
%
\section{Burgers turbulence and instantons} Consider the stochastic Burgers equation
\begin{equation}
\partial_{t} u + u \partial_x u - \partial_{x x}u = \chi^{1/2} * \eta \; ,
\label{eq:burg}
\end{equation}
with periodic boundary conditions in space and Gaussian forcing
\begin{equation}
\left\langle\eta(x,t) \eta(x',t') \right\rangle =\sigma^2  \delta(x - x') \delta(t - t')\,,
\end{equation}
acting only on large length scales through $\chi$. In eq.~(\ref{eq:burg}), the symbol $*$ denotes spatial convolution. For concreteness, we consider the forcing correlation
$\chi(x) = - \partial_{xx}( e^{-x^2/2})$, in a box of size $L_{\mathrm{box}} = 2 \pi$.
In the units of eq.~(\ref{eq:burg}) with viscosity $\nu = 1$, the Reynolds number $\mathrm{Re}$ is varied through the forcing variance $\sigma^2$. All quantities that we are interested in can be written as $\left \langle f[u(\cdot, T)] \right \rangle$ where the ensemble average is with respect to the random forcing realizations, and $T = 1$ large enough that the system is approximately statistically stationary.
The instanton approach is based on a saddle point approximation of the Martin--Siggia--Rose--Janssen--de Dominicis path integral
for averages over eq.~(\ref{eq:burg}). The method consists of three steps~\cite{grafke-grauer-schaefer_2015}: (i) calculation of the instanton configuration, the extremum; (ii) calculation of the contributions from fluctuations around the instanton, a functional determinant at leading order; (iii) contributions of broken continuous symmetries through zero modes if necessary~\cite{schorlepp-grafke-grauer:2023,burekovic-schaefer-grauer:2024}.
In this letter, we compute the one-point PDF of the VG $\rho(a) = \left \langle \delta(\partial_x u(0, T) - a) \right \rangle$ this way at moderate $\mathrm{Re}$ or~$\sigma^2$, and combine the results with fusion rules introduced below.
The instanton approximation to the PDF is exact in the limit $\sigma^2 \to 0$,
when the flow is Gaussian, but also in the limit $| a | \to \infty$ at fixed $\sigma^2$, i.e.\ for high moments of the VG $\langle (\partial_x u)^n \rangle$
with $n \gg 1$ which depend on the tails of the VG PDF.
Instanton configurations for the VG PDF of the Burgers eq.~(\ref{eq:burg}) physically correspond to prototypical ramps ($a>0$) and shocks ($a < 0$).
In particular, while the computation of instantons for the Burgers equation is well-established~\cite{chernykh-stepanov:2001,grafke-grauer-schaefer-etal:2015,simonnet:2023}, a series of recent works~\cite{schorlepp-grafke-grauer:2021,bouchet-reygner:2022,schorlepp-grafke-grauer:2023,schorlepp-tong-grafke-stadler:2023,grafke-schaefer-vanden-eijnden:2023,schorlepp-grafke:2025} has made it possible to evaluate the contribution of Gaussian fluctuations around the instanton in closed form. This step considerably improves the accuracy of the instanton approximation. We follow the Fredholm determinant approach of refs.~\cite{schorlepp-tong-grafke-stadler:2023,schorlepp-grafke:2025} to compute the fluctuations. We briefly present all necessary steps for the theoretical computation of the VG PDF in the following, and refer the reader to ref.~\cite{schorlepp-tong-grafke-stadler:2023} for a more comprehensive, didactic introduction.
For a range of VG values $a$, we approximate the VG PDF to one-loop order around the instanton via
\begin{equation}
    \rho(a) \approx \left(2 \pi \sigma^2\right)^{-1/2} C(a) e^{-S^{\mathrm{I}}(a) / \sigma^2}
    \label{eq:inst-approx-pdf}
\end{equation}
by (i) numerically finding the instanton field $u_a^{\mathrm{I}}(x,t)$ that globally minimizes the 
Onsager--Machlup action
\begin{equation}
\textstyle
    S[u] = \frac{1}{2} \int_0^T \mathrm{d} t \int_0^{L_\mathrm{box}} \mathrm{d} x \left[\partial_t u + u \partial_x u - \partial_{xx} u  \right] \chi^{-1} * [\dots]
\end{equation}
(where $[\dots]$ stands for a repetition of the previous term) under the constraints $u(x, 0) = 0$ and $\partial_x u(0, T) = a$. For eq.~(\ref{eq:burg}) and VGs, there is a unique instanton~$u_a^{\mathrm{I}}$ for each~$a$, and we write $S^{\mathrm{I}}(a) = S[u_a^{\mathrm{I}}]$ for its action, yielding the exponential contribution in eq.~(\ref{eq:inst-approx-pdf}). The instanton $u = u_a^{\mathrm{I}}$, with conjugate momentum density $p = p_a^{\mathrm{I}}$ and Lagrange multiplier ${\cal F}_a$, satisfies the instanton equations~\cite{gurarie-migdal:1996,balkovsky-etal:1997,chernykh-stepanov:2001}
\begin{eqnarray}    \label{eq:inst}
        &\left\{\begin{array}{ll}
        \partial_t u + u \partial_x u - \partial_{xx} u = \chi * p\,,\\
        \partial_t p + u \partial_x p + \partial_{xx} p = 0\,,
        \end{array} \right.\\
        &\mathrm{subject\;to} \left\{\begin{array}{ll}
        u(x,0) = 0\,, \quad \partial_x u(0, T) = a\,,\\
        p(x,T) = -{\cal F}_a \delta'(x)\,,
        \end{array} \right.
\end{eqnarray}
Our numerical, optimal control-based approach to solve this coupled forward-backward problem follows ref.~\cite{schorlepp-grafke-may-grauer:2022}.
(ii) The prefactor $C(a)$, in a Gaussian approximation of fluctuations around the instanton, is given by~\cite{schorlepp-tong-grafke-stadler:2023,schorlepp-grafke:2025}
\begin{equation}
C(a) = | {\cal F}_a | /  \left(2 S^{\mathrm{I}}(a) \det \left(\mathrm{Id} -
B_a \right) \right)^{1/2}\,.
\label{eq:fredh}
\end{equation}
Here, the integral operator $B_a$ corresponds to the second variation of the noise-to-observable map at the instanton, and $\det$ is a Fredholm determinant which can be computed iteratively in a matrix-free way, by finding the largest eigenvalues of $B_a$~\cite{schorlepp-tong-grafke-stadler:2023,schorlepp-grafke:2025}. For this, we only need to apply the operator $B_a$ to different test noise perturbations $\delta \eta(x,t)$: We have $B_a \delta \eta = {\cal P} \left[\chi^{1/2} * \delta p\right]$ where $\delta p$ is found through solving~\cite{schorlepp-tong-grafke-stadler:2023,schorlepp-grafke:2025}
\begin{eqnarray} \label{eq:scd-var}
        &\left\{\begin{array}{ll}
        \partial_t \delta u +\partial_x \left(u_a^{\mathrm{I}} \delta u \right) - \partial_{xx} \delta u = \chi^{1/2} * {\cal P} \left[\delta \eta \right]\,,\\
        \partial_t \delta p + u_a^{\mathrm{I}} \partial_x \delta p + \delta u \partial_x p_a^{\mathrm{I}} + \partial_{xx} \delta p = 0\,,
        \end{array} \right.\\
        &\mathrm{with}\left\{\begin{array}{ll}
        \delta u(x,0) = 0\,,\\
        \delta p(x,T) = 0\,,
         \end{array} \right.
\end{eqnarray}
via one forward-backward pass, with ${\cal P}$ the $L^2$-orthogonal projection onto $\left( \chi^{1/2} * p_a^{\mathrm{I}} \right)^\perp$.
As in ref.~\cite{schorlepp-kormann-luebke-etal:2025}, we use a pseudo-spectral spatial discretization of all equations with resolutions  $n_x \in \{512, 2048\}$, as well as the Heun scheme with integrating factor for the time stepping with $n_t = 10^4$ equidistant time steps for the instanton computations. Instantons and prefactors were computed for approximately $500$ values of $a \in [-30000, -0.1]$ and $70$ values of $a \in [0.1, 223]$ (since the positive tail is strongly suppressed), both logarithmically spaced, with $200$ dominant eigenvalues of $B_a$ per $a$ to approximate the Fredholm determinant~(\ref{eq:fredh}). This computation needs to be done only once and can then be rescaled to different $\sigma^2$ and hence Re via eq.~(\ref{eq:inst-approx-pdf}). In other words: the instanton action $S^{\mathrm{I}}(a)$ and fluctuation prefactor $C(a)$ in eq.~(\ref{eq:inst-approx-pdf}) do not depend on $\sigma^2$ and hence $\mathrm{Re}$ explicitly -- as $\sigma^2$ is varied their \textit{relative weight} at given gradient strengths $a$ in the VG PDF shifts:
\begin{equation}
    \rho(a) \approx e^{-\frac{1}{2} \ln(2 \pi \sigma^2 ) - \frac{1}{\sigma^2} \left( S^{\mathrm{I}}(a) - \sigma^2 \ln C(a) \right)}\,.
\end{equation}
In addition to the instanton computations, we also perform $10^4$ independent DNS runs of eq.~(\ref{eq:burg}) until $T=1$ for different forcing variances $\sigma^2 \in [10^{-2}, 3 \cdot 10^5]$ with $n_x = 2048$, both to have data to compare our instanton predictions to, as well as to obtain the Taylor--Reynolds number $\mathrm{Re}_\lambda = \lambda u_{\mathrm{rms}} / \nu$ (where $u_{\mathrm{rms}} = \sqrt{\left \langle u^2 \right \rangle}$, and $\lambda = u_{\mathrm{rms}} \sqrt{2 \nu / \left \langle \varepsilon\right \rangle}$ with $\varepsilon = 2 \nu \left(\partial_x u\right)^2$, for $\nu = 1$ here) as a function of~$\sigma^2$, as shown in fig.~\ref{fig:re-sig-dependence}. For small~$\sigma^2$, when the flow is approximately Gaussian, we have $\mathrm{Re}_\lambda \propto \sigma$, and for large~$\sigma^2$ this becomes $\mathrm{Re}_\lambda \propto \mathrm{Re}^{1/2}  \propto \sigma^{1/3}$~\cite{grafke-grauer-schaefer-etal:2015,schorlepp-grafke-grauer:2021,schorlepp-grafke-may-grauer:2022}.
%
%
\section{Fusion rules}
Fusion rules~\cite{lvov-procaccia:1996, benzi-biferale-toschi:1998, schumacher-sreenivasan-yakhot:2007} allow us to relate scaling laws of ``fused'' observables, such as gradients,
to the Re scaling of 
multi-point observables as those points collapse.
Since the Burgers eq.~(\ref{eq:burg}) is pressureless, we apply the fusion rules of ref.~\cite{benzi-biferale-toschi:1998}, consistent with Nelkin scaling~\cite{nelkin:1990}.
We consider the standardized even-order VG moments given by eq.~(\ref{eq:mn-def}). At low Re, $u$ and its derivatives
are Gaussian, hence
$
M_n(\mathrm{Re}_\lambda \ll 1) \approx (n-1)!! 
$.
At $\mathrm{Re}_\lambda \approx 1$, there is a transition to turbulence, with
the theoretical result at large Re~\cite{chakraborty-frisch-pauls-ray:2012, friedrich-margzoglou-biferale-etal:2018}
\begin{equation}
M_n(\mathrm{Re}_\lambda \gg 1) \propto \mathrm{Re}_\lambda^{n-2}\,.
\label{eq:mn-theory}
\end{equation}
Remarkably, the standardized
VG moments were shown to display this scaling already
at moderate $\mathrm{Re}_\lambda \gsim 1$
in DNS~\cite{chakraborty-frisch-pauls-ray:2012,friedrich-margzoglou-biferale-etal:2018}; see ref.~\cite{chakraborty-frisch-pauls-ray:2012} for an explanation via subleading corrections to~(\ref{eq:mn-theory}) from the Cole--Hopf solution of the Burgers equation.
The exponents $\zeta_q$ of structure functions
\begin{equation}
    S_q(r) = \left \langle | u(x+r) -u(x)|^q \right \rangle \propto r^{\zeta_q}
\end{equation}
for $r$ in the inertial range at large Re are well-known: 
\begin{equation}
    \zeta_q = \min \{1,q\}
    \label{eq:zeta-q-theory}
\end{equation}
for the Burgers eq.~(\ref{eq:burg}).
In general, the fusion rules derived in ref.~\cite{benzi-biferale-toschi:1998} (see also ref.~\cite{friedrich-margzoglou-biferale-etal:2018} for further discussion) imply
\begin{equation}
    M_n(\mathrm{Re}_\lambda \gg  1) \propto \mathrm{Re}_\lambda^{\theta_n}\,,
\end{equation}
where the exponents $\theta_n = 2 \phi_{n}-n \phi_{2}$ encode the structure function scaling via
$
    \phi_{n}=\frac{1}{2}\left(q-\zeta_{q}\right)
$.
Here, 
$q = q(n)$ is determined through $n = \frac{1}{2} (q + \zeta_q)$, with $\phi_n$ being the scaling exponents for moments of the VG 
$
\langle (\partial_x u)^n \rangle \propto \mathrm{Re}^{\phi_n} \propto \mathrm{Re}_\lambda^{2 \phi_n}
$.
This indeed yields eq.~(\ref{eq:mn-theory}) when starting from eq.~(\ref{eq:zeta-q-theory}), with $\phi_n = n - 1$.
Conversely, given pairs $(n, \theta_n)$, which we find using instanton calculus, we need to invert this construction to obtain $q \mapsto \zeta_q$. We fix the exact relation $\zeta_3 = 1$~\cite{frisch:1995}, such that $\phi_2 = 1$, to uniquely determine $\phi_n$ from $\theta_n$ for even integers $n \geq 2$. To numerically find $q \mapsto \zeta_q$ from this, one can interpolate the obtained $(n, \theta_n)$, or prescribe a parametric form of $q \mapsto \zeta_q$. We follow the latter strategy below.
To summarize, we have proposed to first map the velocity gradient moments $M_n$ to their scaling exponents $\theta_n$, which fusion rules then relate to $\zeta_q$.
\begin{table}
    \caption{Scaling exponents $\theta_n$ of the normalized VG moments~(\ref{eq:mn-def}), as determined from fig.~\ref{fig:mn-result} (dashed lines). The reported fit errors are the standard error of the estimated slope of $(\ln \mathrm{Re}_\lambda, \ln M_n)$ assuming residual normality.}
    \label{tab:vg-expos}
    \centering
    \begin{tabular}{cccc}
        $n$ & $\theta_n$ (computed) & $\theta_n$ (theory) & rel.\ error (\%)\\[2pt]
         \hline\\[-10pt]
        $4$  & $1.83 \pm 0.05$ & $2$  & $-8.6 \pm 2.6$\\
        $6$  & $3.47 \pm 0.05$ & $4$  & $-13.3 \pm 1.2$\\
        $8$  & $5.08 \pm 0.06$ & $6$  & $-15.3 \pm 0.9$\\
        $10$ & $6.68 \pm 0.07$ & $8$  & $-16.4 \pm 0.9$\\
        $12$ & $8.28 \pm 0.09$ & $10$ & $-17.2 \pm 0.9$
    \end{tabular}
\end{table}
%
%
\begin{figure}
    \centering
    \includegraphics[width=.7\linewidth]{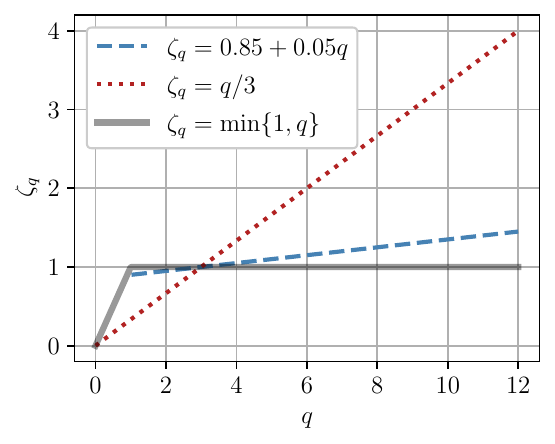}
    \caption{Exact structure function exponents $\zeta_q$ (grey, solid) for the Burgers eq.~(\ref{eq:burg}), K41 theory prediction (red, dotted), and exponents from instantons and fusion rules (blue, dashed).}
    \label{fig:exponents}
\end{figure}
\section{Results} 
\begin{figure}
    \centering
    \includegraphics[width=.85\linewidth]{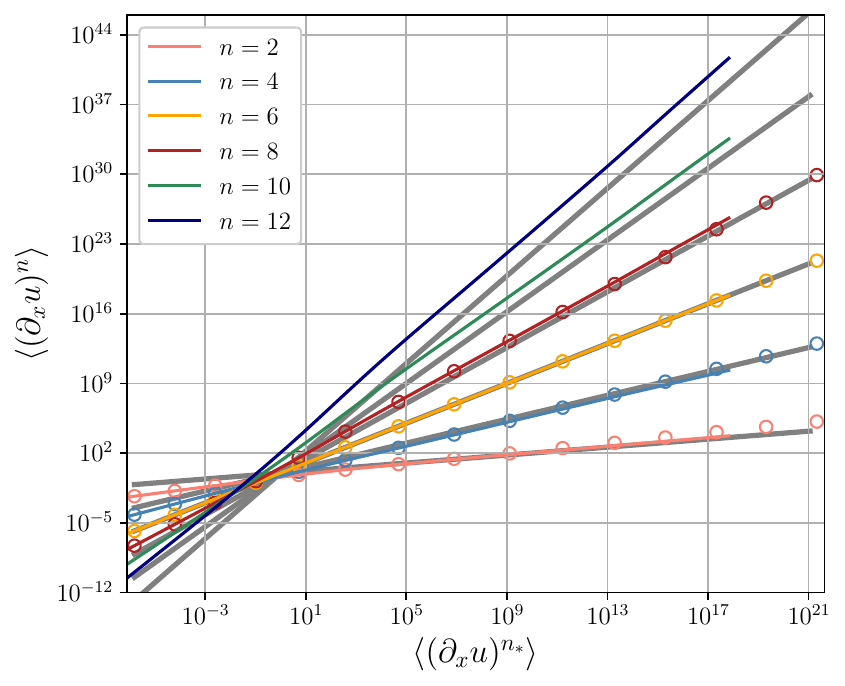}
    \caption{ESS plot for VG moments. Solid colored lines show the $n$-th VG moment as a function of the $(n_* = 6)$-th one, obtained from integrating over the one-loop VG PDF~(\ref{eq:inst-approx-pdf}) at different $\sigma^2$. The circles show DNS results, and grey lines the theoretically expected exponent $\frac{n-1}{n_* - 1}$ as their slope.}
    \label{fig:ess-ux}
\end{figure}

We show the obtained VG moments from the instanton calculations, compared to DNS and known theoretical results, in fig.~\ref{fig:mn-result}. $\mathrm{Re}_\lambda$ on the horizontal axis is found from DNS. VG moments from DNS for $n>8$ could not be estimated since more tail data would be required.
In contrast, the instanton approximation~(\ref{eq:inst-approx-pdf}) can be evaluated arbitrarily far into the PDF tails.
The DNS results show a clear crossover at $\mathrm{Re}_\lambda \approx 1$ to the scaling~(\ref{eq:mn-theory})~\cite{friedrich-margzoglou-biferale-etal:2018}.
Without the one-loop prefactor $C(a)$ in eq.~(\ref{eq:inst-approx-pdf}), and integrating only $\rho(a) \approx \mathrm{const} \times \exp \left(-S^{\mathrm{I}}(a) / \sigma^2\right)$ for the numerator and denominator of eq.~(\ref{eq:mn-def}), the bare instanton fails to capture this transition (dotted lines in fig.~\ref{fig:mn-result}; with const.\ chosen to normalize the PDF to integral 1).
By including the prefactor $C(a)$, i.e., integrating eq.~(\ref{eq:inst-approx-pdf}) for numerator and denominator moments in eq.~(\ref{eq:mn-def}), the scaling crossover at $\mathrm{Re}_\lambda \approx 1$ is correctly captured (solid lines in fig.~\ref{fig:mn-result}).
Hence, accounting for fluctuations around instantons is crucial here.
Still, no clean power law scaling at $\mathrm{Re}_\lambda \gsim  1$ is visible, and the slope of the predicted $M_n$ is too steep compared to the theoretically known result eq.~(\ref{eq:mn-theory}).
In order to understand this discrepancy, we show the moment integrand $a^n \rho(a)$ for different $n$ in fig.~\ref{fig:moment-integrands}.
This confirms that eq.~(\ref{eq:inst-approx-pdf}) becomes more accurate as $n$ increases, but a constant overall factor is missed by the PDF approximation~(\ref{eq:inst-approx-pdf}) (while the PDF tail scaling is already well captured~\cite{schorlepp-kormann-luebke-etal:2025}).
As is well known, this factor stems from the fact that the approximation deviates from the true PDF in the core of the distribution where the saddle-point dominance breaks down, thereby producing an incorrect overall normalization.
Motivated by this insight, we rescale the one-loop instanton PDF~(\ref{eq:inst-approx-pdf}) by a constant, $\mathrm{Re}$-dependent factor for all $a < 0$ (since the right tail is exponentially  suppressed already at moderate Re) to match the left tails of the respective DNS PDF, cf.\ table~\ref{tab:params}.
Furthermore, we use DNS instead of instantons to evaluate the denominator in eq.~(\ref{eq:mn-def}), because the second moment is not a tail quantity and cannot be accurately captured by the instanton approximation in its current form (cf.\ fig.~\ref{fig:moment-integrands}).
Note that these DNS inputs (tail rescaling, second moment, and the relation between~$\sigma^2$ and~$\mathrm{Re}_\lambda$) are used in this letter as a hybrid approach based on instantons, DNS and fusion rules -- alternatively, it delineates the predictive power of the instanton approach for high moments, provided these low-order DNS statistics \textit{could} be perfectly captured by a theoretical method.
These modifications lead to the dashed lines in fig.~\ref{fig:mn-result},
which display clean power law scaling for all shown moments and excellently match DNS data where available. Linear regression on the obtained $(\ln \mathrm{Re}_\lambda, \ln M_n)$ prediction for $2 \leq \mathrm{Re}_\lambda \leq 15$ yields the scaling exponents $\theta_n$ of the normalized VG moments listed in table~\ref{tab:vg-expos}.
The remaining errors can clearly be attributed to ``finite-size effects'', i.e.\ the small $\mathrm{Re}_\lambda$ considered here, and are visible already in the DNS data in figs.~\ref{fig:mn-result} and~\ref{fig:re-sig-dependence}, where they begin to disappear as $\mathrm{Re}_\lambda$ increases.
In order to infer $q \mapsto \zeta_q$ from the $(n, \theta_n)$ pairs in Table~\ref{tab:vg-expos}, given the almost perfectly linear behavior of the data and expected monofractality of Burgers turbulence for $q \geq 1$, we suppose $\zeta_q = \alpha + \frac{1-\alpha}{3} q$ with $\alpha$ to be determined ($\beta$-model). The slope of $n \mapsto \theta_n$ from table~\ref{tab:vg-expos} is then consistent with $\zeta_q \approx 0.85 + 0.05 q$.
Although this prediction for the scaling exponents does not exactly reproduce the known large $\mathrm{Re}$ result $\zeta_q = 1$ for $q \geq 1$, cf.\ fig.~\ref{fig:exponents}, because of finite-size effects, it is important to emphasize that it is obtained from a controllable method applied to the evolution equations, and extends to large $q$.
%
%
To demonstrate that the core predictive power of the instanton approach is independent of the low-order DNS inputs, we apply ideas from ESS~\cite{benzi-ciliberto-tripiccione:1993,benzi-ciliberto-baudet:1993,chakraborty-frisch-ray:2010,chakraborty-frisch-pauls-ray:2012} and compare only high VG moments -- avoiding input from DNS altogether since not even $\mathrm{Re}_\lambda$ as a function of $\sigma^2$ is needed. As an example, we show $\left \langle (\partial_x u)^n \right \rangle$ as a function of $\left \langle (\partial_x u)^{n_*} \right \rangle$ in fig.~\ref{fig:ess-ux} for $n_* = 6$. Notably, the solid colored lines are produced directly from eq.~(\ref{eq:inst-approx-pdf}) without any DNS input for rescaling or low moments. As expected, we see that focusing on high, relative moments gives excellent agreement with the theory and DNS and thus demonstrates the internal consistency of the instanton-based approach and its capability to predict high-order structure function scaling.
\section{Conclusion and outlook}
In this letter, we have introduced a hybrid
instanton–fusion-rule framework, combining the instanton approach at the onset of turbulence with fusion rules and DNS to infer structure function exponents in fully developed Burgers turbulence.
This approach offers a systematic framework that can capture the turbulent regime, and specifically targets high structure functions via the instanton approach, which are otherwise difficult to estimate from DNS.
The methodology could be applied to other systems such as the 3D NSE, passive scalar turbulence, or MHD turbulence in the future.
Our strategy has been to use instanton calculus where it gives accurate results -- at moderate Re, and for dissipative VG tail events -- and combine it with additional input -- fusion rules and DNS -- to ``upscale''  to the inertial range in fully developed turbulence.
A particularly important future direction is the application of this approach to the 3D NSE, where, in contrast to the Burgers test case, it is still an open question whether scaling exponents saturate (as in the Yakhot model~\cite{yakhot:2001,yakhot:2006}) or grow asymptotically linearly with the order of the structure function (as in the She--Leveque model~\cite{she-leveque:1994}).
While computationally more challenging, and requiring the inclusion of zero modes~\cite{schorlepp-grafke-grauer:2023,burekovic-schaefer-grauer:2024} and multiple instanton branches~\cite{schorlepp-grafke-may-grauer:2022}, such an extension to the 3D NSE is feasible~\cite{schorlepp-grafke-may-grauer:2022,schorlepp-tong-grafke-stadler:2023} and might produce structure function exponents in even closer agreement with experimental and DNS data since intermittency is reduced compared to Burgers turbulence.
Still, a possible shortcoming of the method presented here, already for the Burgers test case, is the need for information from DNS for low-order quantities, which fall outside the applicability of the instanton approach. To demonstrate the method's fundamental validity independent of empirical tuning, we used an ESS approach here (see also the recent discussion in ref.~\cite{buaria-pumir:2026}).
Several other strategies are under consideration for future work. A natural extension of the present approach is to include not only quadratic, but also higher-order perturbations around the instanton. 
This would lead to evolution equations for higher-order ($d \geq 3$) tensors compared to the Riccati equation of ref.~\cite{schorlepp-grafke-grauer:2021} of dimensionality $N^d$, where $N$ denotes the number of degrees of freedom (mesh points or Fourier modes), thus requiring additional low-rank techniques~\cite{breiten-dolgov-stoll:2021,abram-venturi:2024}. 
An even more challenging direction is to study the stochastic equation around the instanton~\cite{ebener-etal:2019} using methods from functional renormalization~\cite{canet:2022}. First steps in this direction have recently been presented in refs.~\cite{ihssen-pawlowski:2025,bonanno-ihssen-pawlowski:2025,yang-darcy-hudes-etal:2025}.
Arguably the central challenge is to identify which fluctuations provide the leading contributions to the path integral, potentially those associated with slightly broken zero modes. Establishing this understanding may represent a decisive step toward a theory of turbulence.
%

\acknowledgments RG acknowledges support from
the DFG via SFB1491.

\bibliographystyle{eplbib}
\bibliography{literature}


\section{Appendix}

\begin{figure}[h]
    \centering
    \includegraphics[width = .35 \textwidth]{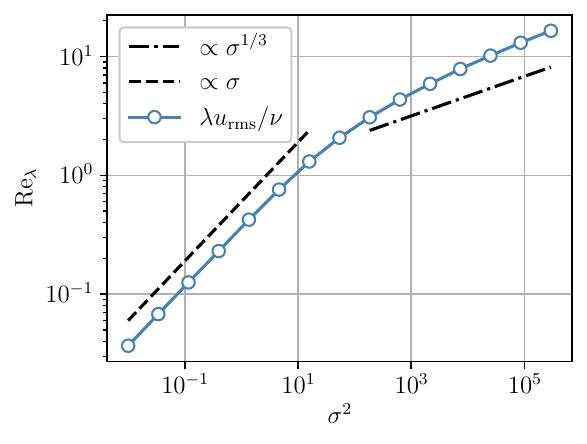}
    \caption{Taylor--Reynolds number $\mathrm{Re}_\lambda = \lambda u_{\mathrm{rms}} / \nu$ as a function of the forcing strength $\sigma^2$ from DNS of eq.~(\ref{eq:burg}).}
    \label{fig:re-sig-dependence}
\end{figure}

\begin{figure*}
    \centering
    \includegraphics[width=.472\textwidth]{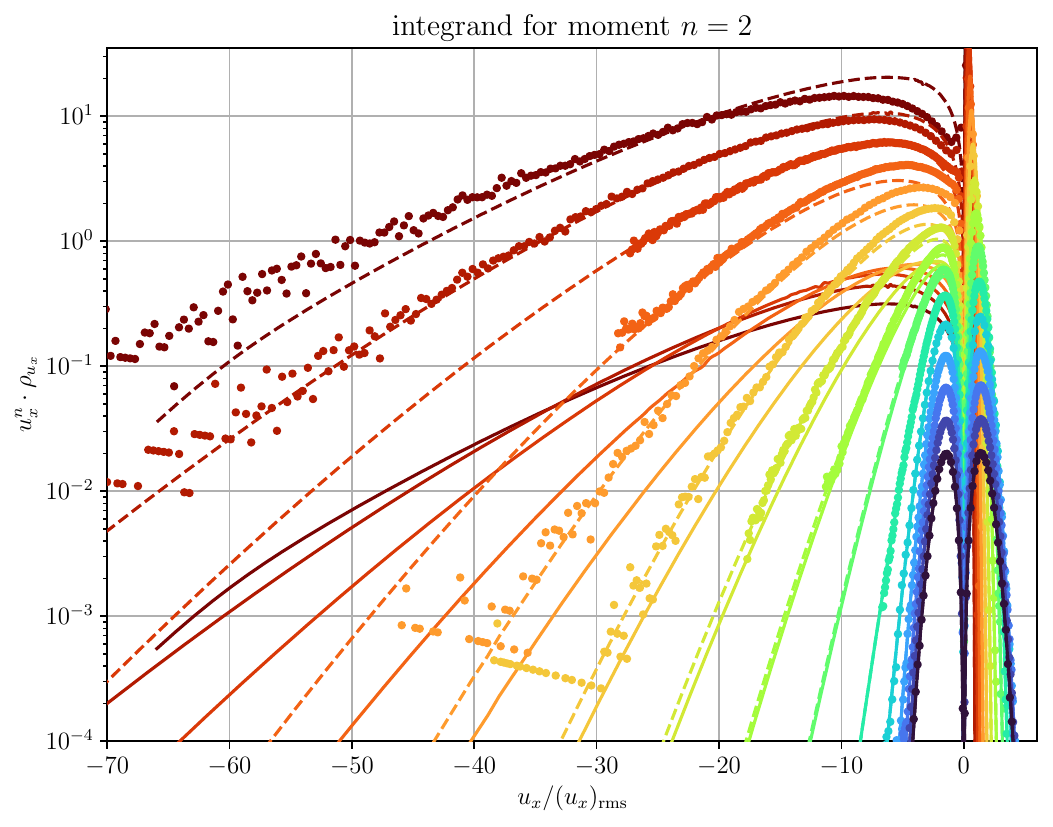} \hfill
    \includegraphics[width=.472\textwidth]{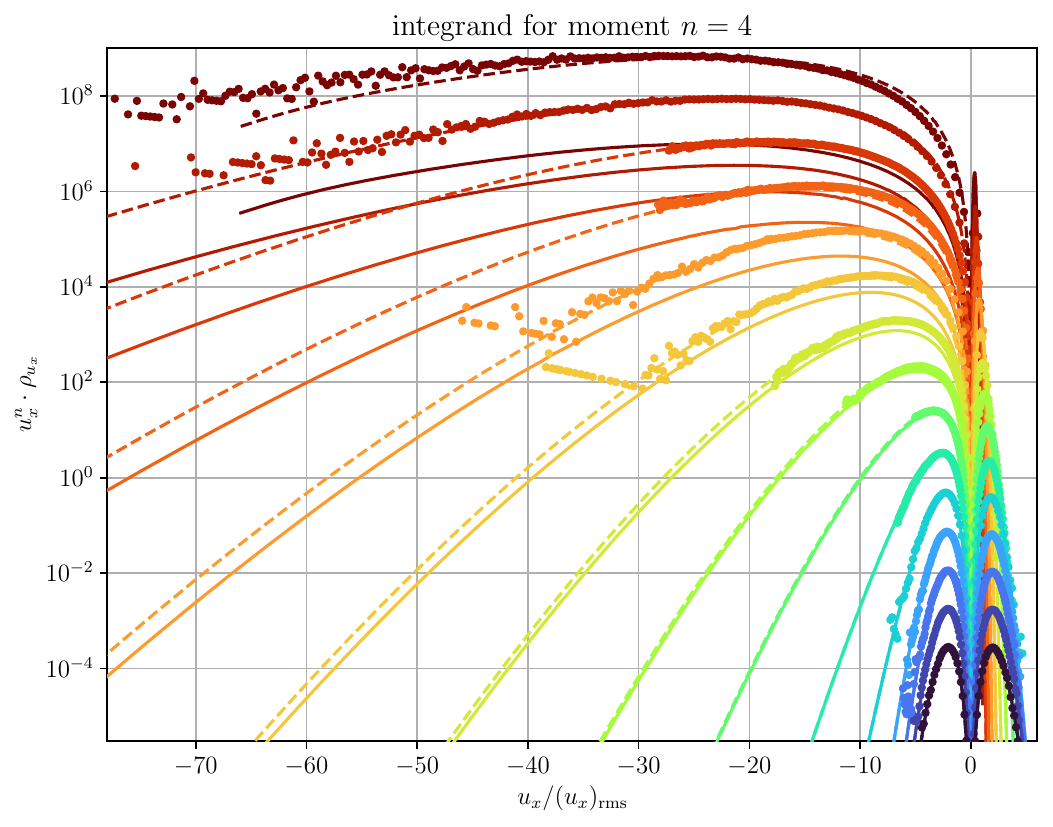} \\
    \includegraphics[width=.472\textwidth]{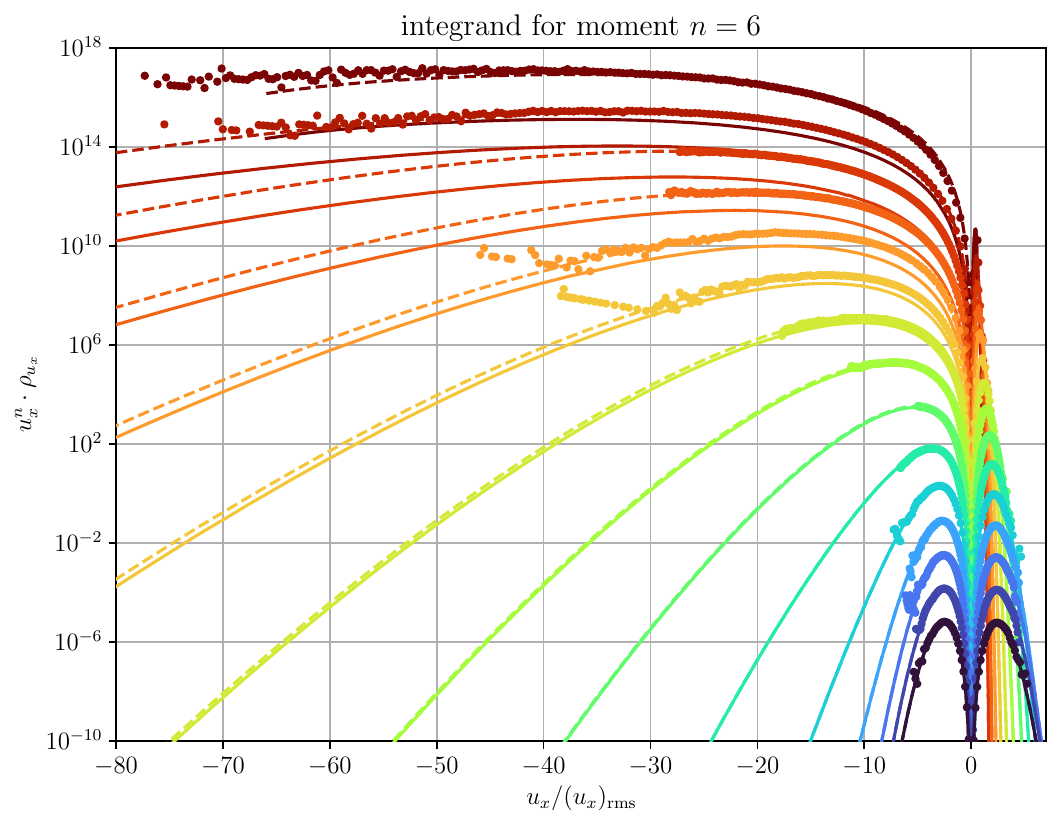} \hfill
    \includegraphics[width=.472\textwidth]{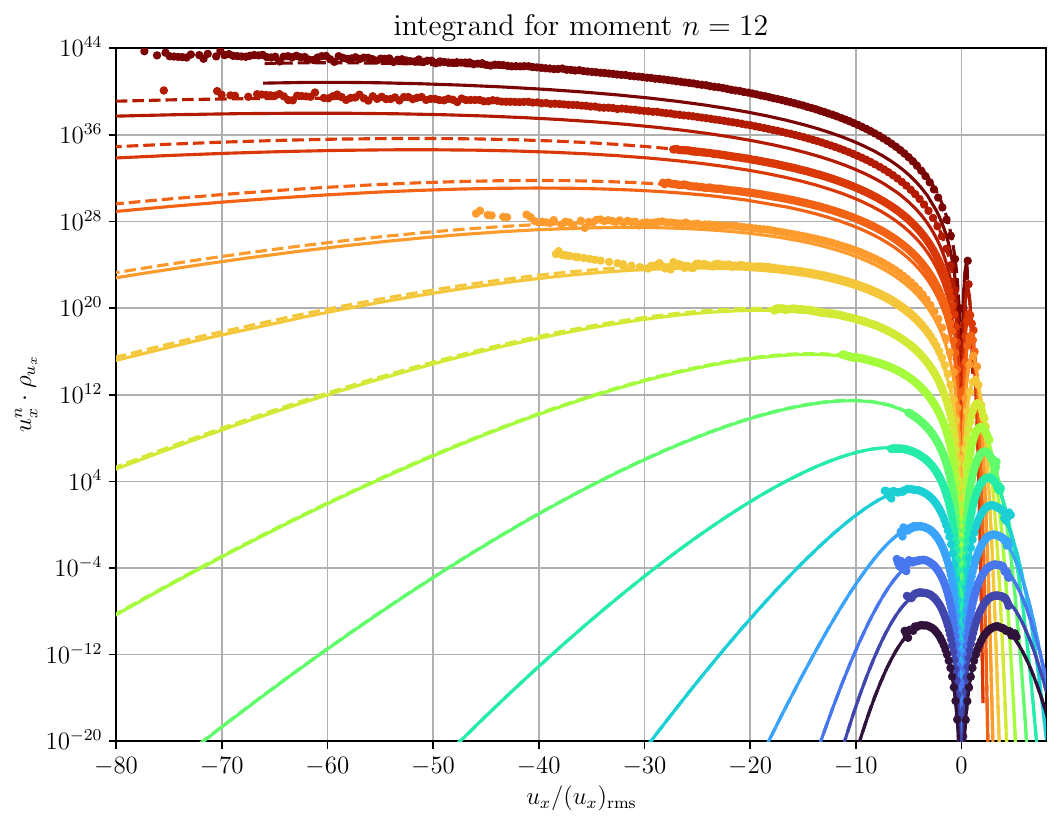} 
    \caption{Moment integrands $a^n \rho(a)$ of the VG PDF. Dots show DNS data, solid lines show eq.~(\ref{eq:inst-approx-pdf}), and dashed lines are the same curves multiplied by a constant (cf.\ table~\ref{tab:params}) for $a < 0$ to match DNS PDF tails. See table~\ref{tab:params} for noise variances $\sigma^2$ and Taylor--Reynolds numbers $\mathrm{Re}_\lambda$ that different colors correspond to. Note that bin ranges for DNS PDFs were set automatically and do not capture the full range of available data for some $\mathrm{Re}_\lambda$. This does not affect other results, as DNS VG moments were computed directly from field data without binning.}
    \label{fig:moment-integrands}
\end{figure*}

\noindent This section contains additional figures and tables: fig.~\ref{fig:re-sig-dependence} shows
$\mathrm{Re}_\lambda$ as a function of the forcing strength $\sigma^2$ in DNS,
fig.~\ref{fig:moment-integrands} shows the integrands $a^n \rho(a)$
for the VG moments~(\ref{eq:mn-def}),
and table~\ref{tab:params} lists parameters
and rescaling constants to match DNS PDF tails with eq.~(\ref{eq:inst-approx-pdf}).

\begin{table*}
    \centering
    \caption{Noise variances $\sigma^2$ and Taylor-Reynolds numbers $\mathrm{Re}_\lambda$ in DNS of eq.~(\ref{eq:burg}), and normalization of one-loop PDFs~(\ref{eq:inst-approx-pdf}) before and after rescaling the left tail by the listed constant to match DNS PDF tails.}
    \label{tab:params}
    \begin{tabular}{cccccc}
         $\sigma^2$ & $\mathrm{Re}_\lambda$ & color in fig.~\ref{fig:moment-integrands} &  $\int_{-\infty}^\infty$ eq.~(\ref{eq:inst-approx-pdf}) $\mathrm{d} a$ & left tail rescale & $\int_{-\infty}^\infty$ eq.~(\ref{eq:inst-approx-pdf}) $\mathrm{d} a$ after rescaling\\[2pt]
         \hline\\[-10pt]
 0.010 & 0.04 & \textcolor[HTML]{30123b}{$\bullet$} & 0.999935 & 1 & 0.999935 \\
 0.034 & 0.07 & \textcolor[HTML]{4146ac}{$\bullet$} & 0.999822 & 1 & 0.999822 \\
 0.117 & 0.13 & \textcolor[HTML]{4776ee}{$\bullet$} & 0.999473 & 1 & 0.999473 \\
 0.398 & 0.23 & \textcolor[HTML]{3aa3fc}{$\bullet$} & 0.998309 & 1 & 0.998309 \\
 1.359 & 0.42 & \textcolor[HTML]{1bd0d5}{$\bullet$} & 0.994445 & 1 & 0.994445 \\
 4.642 & 0.76 & \textcolor[HTML]{25eca7}{$\bullet$} & 0.982510 & 1 & 0.982510 \\
15.849 & 1.31 & \textcolor[HTML]{61fc6c}{$\bullet$} & 0.952587 & 1.1 & 0.992669 \\
54.117 & 2.07 & \textcolor[HTML]{a4fc3c}{$\bullet$} & 0.901730 & 1.2 & 0.965919 \\
184.785 & 3.08 & \textcolor[HTML]{d2e935}{$\bullet$} & 0.850983 & 1.5 & 0.964793 \\
630.957 & 4.35 & \textcolor[HTML]{f4c73a}{$\bullet$} & 0.832163 & 2   & 0.978337  \\
2154.435 & 5.90 & \textcolor[HTML]{fe9b2d}{$\bullet$} & 0.864512 & 3   & 1.040367 \\
7356.423 & 7.86 & \textcolor[HTML]{f36315}{$\bullet$} &0.954953 & 5   & 1.158280  \\
25118.864 & 10.17& \textcolor[HTML]{da3907}{$\bullet$} &1.109413 & 11  & 1.396104 \\
85769.590 & 13.06& \textcolor[HTML]{b21a01}{$\bullet$} &1.346033 & 24  & 1.711882 \\
292864.456 &16.48& \textcolor[HTML]{7a0403}{$\bullet$} &1.731763 & 65  & 2.290340 \\
    \end{tabular}
\end{table*}

\end{document}